\documentclass[preprint,floatfix,longbibliography,superscriptaddress]{revtex4-1} 
\usepackage{amsmath}  % needed for \tfrac, \bmatrix, etc.
\usepackage{amsfonts} % needed for bold Greek, Fraktur, and blackboard bold
\usepackage{graphicx} % needed for figures
\usepackage{bm}
\usepackage[usenames,dvipsnames]{color}
\usepackage{verbatim}
\usepackage[T1]{fontenc}
\usepackage{MnSymbol}
\usepackage{wasysym}
\usepackage{soul}
\usepackage{anyfontsize}
\usepackage{subfig} % added by Susan to do multi-part figures 
\usepackage{ upgreek }

\def\be{\begin{eqnarray}}
\def\ee{\end{eqnarray}}

\def\He3{$^3$He}

\def\He3{$^3$He}

\RequirePackage{lineno}

\begin{document} 

\setpagewiselinenumbers

\modulolinenumbers[5]

\title{A Synchronous Spin-Exchange Optically Pumped NMR-Gyroscope}
\def\wisc{Department of Physics, University of Wisconsin-Madison, Madison, Wisconsin 53706, USA}
\def\sand{Sandia National Laboratories, Albuqerque, NM 87185, USA.}
\author{Susan S. Sorensen}
\affiliation{\wisc}
\author{Daniel A. Thrasher}
\affiliation{\sand}
\author{Thad G. Walker}
\affiliation{\wisc}
\email{tgwalker@wisc.edu}

\date{\today}

\begin{abstract}{
Inertial navigation systems generally consist of timing, acceleration, and orientation measurement units. Although much progress has been made towards developing primary timing sources such as atomic clocks, acceleration and orientation measurement units often require~calibration. Nuclear Magnetic Resonance (NMR) gyroscopes,  which rely on continuous measurement of the simultaneous Larmor precession of two co-located polarized noble gases,  can be configured to have scale factors that depend to first order only on fundamental constants. The noble gases are polarized by spin-exchange collisions with co-located optically pumped alkali-metal atoms. The alkali-metal atoms are also used to detect the phase of precession of the polarized noble gas nuclei. Here we present a~version of an NMR gyroscope designed to suppress systematic errors from the alkali-metal~atoms. We~demonstrate rotation rate angle random walk (ARW) sensitivity of $16\; \upmu \text{Hz}/\sqrt{\text{Hz}}$ and bias instability of $\sim$800 nHz.
}\end{abstract}

\maketitle

\section{Introduction} \label{intro}
%The introduction should briefly place the study in a broad context and highlight why it is important. It should define the purpose of the work and its significance. The current state of the research field should be reviewed carefully and key publications cited. Please highlight controversial and diverging hypotheses when necessary. Finally, briefly mention the main aim of the work and highlight the principal conclusions. As far as possible, please keep the introduction comprehensible to scientists outside your particular field of research. Citing a journal paper \cite{ref-journal}. And now citing a book reference \cite{ref-book}. Please use the command \citep{ref-journal} for the following MDPI journals, which use author-date citation: Administrative Sciences, Arts, Econometrics, Economies, Genealogy, Humanities, IJFS, JRFM, Languages, Laws, Religions, Risks, Social Sciences.

%\textls[-25]{
Spin-exchange pumped NMR gyroscopes~\cite{Thrasher2019,thrasher2019PRApp,Walker2016,Kornack2005,Jiang2018,Karwacki1980} use the precession of spin-polarized nuclei to measure~rotation.  
%consist of 
A vapor cell contains one or more isotopes of noble gas atoms and an~alkali-metal~vapor. The alkali-metal vapor is optically pumped using circularly polarized near-infrared laser light. Spin-exchange collisions between the polarized alkali-metal and noble gas atoms polarize the noble gas nuclei.~When subject to a magnetic (or ``bias'') field, the polarized species undergo Larmor precession. By~simultaneously measuring the Larmor precession of each entity, magnetic field correlations can be removed, thereby increasing sensitivity to non-magnetic spin-dependent phenomena, such as inertial~rotation.%}

%\textls[-5]{
The precession of a polarized gas about a constant bias field constitutes an inertial reference frame, i.e., the gas has (ideally) no way of knowing whether or not its container is rotating. If the precession of the polarized gas is measured relative to a fixed point in the laboratory frame, such as with a pick-up~coil, then rotation of the pick-up coil about the bias field will change the measured rate of precession of the polarized gas.%}

The application of polarized gases as inertial sensors is promising because of their low intrinsic~noise, miniaturize-ability, low power consumption, insensitivity to acceleration, and intrinsic scaling from experimental observable to rotation that is independent of experimental parameters~\cite{Donley2013,Brinkmann62}. Interest in novel navigation systems which operate in global positioning system (GPS) denied environments has increased as of late. To date, the United States, Russia, China, and India have successfully demonstrated the capability to destroy their own satellites~\cite{Chaudhury2019}. Such capability could potentially be used to remove GPS satellites. Other GPS denied environments include subterranean and submarine travel. One obstacle to harnessing geothermal energy resources is the limited ``in-hole'' navigation of horizontal drilling rigs. The future development of autonomous vehicles also relies on robust navigation, including in GPS denied environments. NMR gyroscopes will likely first be used to provide long-term corrections to another miniaturized gyroscope whose short term noise performance is superior but whose long term drift is inferior to the NMR gyroscope. A proof-of-concept experiment along these lines was recently performed using a classical and quantum accelerometer~\cite{Cheiney2018}. 

Important sources of systematic errors in spin-exchange pumped NMR gyroscopes are longitudinal (i.e., parallel to the bias field) spin-exchange fields produced by polarized alkali-metal atoms or noble gas nuclei~\cite{Thrasher2019,Bulatowicz2013,Walker2016}. Such longitudinal spin-exchange fields are not identical for all noble gas species, and therefore cannot be removed in the same way as classical magnetic field correlations. Longitudinal spin-exchange fields are well suppressed by synchronous spin-exchange optical pumping~\cite{Thrasher2019,Korver2015}, wherein the alkali-metal vapor is optically pumped transverse to a low duty cycle pulsed bias field. Each bias field pulse produces $2\pi$ precession of the alkali-metal atoms such~that, despite having a gyromagnetic ratio in excess of $10^3$x that of the noble gas nuclei, the alkali-metal atoms can effectively co-precess with the noble gas nuclei. 

This manuscript describes a $^{131}$Xe-$^{129}$Xe synchronous spin-exchange optically pumped NMR gyroscope which uses modulation of the alkali-metal ($^{85}$Rb) polarization to drive both Xe isotopes' NMR simultaneously. The Rb is also used as an embedded magnetometer to detect the Xe precession. Compared to our recent work~\cite{Thrasher2019,thrasher2019PRApp} describing NMR excitation by modulating the bias field, polarization modulation (PM) further suppresses the influence of time-averaged longitudinal Rb spin-exchange fields~\cite{Korver2015} by moving such fields from DC to AC. We study how modulation of the embedded Rb magnetometer causes signal mixing on our detection which leads to an effective change in scale factor~\cite{ThrasherThesis}.

\subsection{Bloch Equation} \label{BEsec}
The spin dynamics of the polarized noble gas transverse ($K_+ = K_x+iK_y$) and longitudinal ($K_z$) spins are described by the Bloch equations 
%All variables should be italic. Please carefully check all the variables and parameters in the manuscript and keep them in the same format, italic or not.
\begin{subequations}
\begin{eqnarray}
{dK_+\over dt}=-(\mp i\Omega_z^K+\Gamma_2)K_+ +\Gamma_S^K S_+ + \mp i\Omega_+^K K_z,\label{BE1}
\\
{dK_z\over dt}=\mp (\Omega_y^K K_x-\Omega_x^K K_y)-\Gamma_1 K_z,\label{BE2}
\end{eqnarray}
\end{subequations}
where $\mathbf{\Omega}^K$ is the Xe resonance frequency arising from both the Larmor precession $\gamma^K \mathbf{B}$ and from rotation $\omega^R \mathbf{\hat{z}}$, $\Gamma_1$ ($\Gamma_2$) is the longitudinal (transverse) relaxation rate, $\mathbf{S}$ is the Rb polarization, and~$\Gamma_S^K$ is the spin-exchange rate constant. Let superscript $a$ represent $^{129}$Xe and superscript $b$ represent $^{131}$Xe. Since $\gamma^a <0$ and $\gamma^b > 0$ we find it useful to write the gyromagnetic ratio $\gamma$ (and hence $\Omega$) as a~positive~value. The sign is written explicitly out front (top sign is for $a$, bottom sign is for $b$). 

We null the transverse fields experienced by the Xe, including the spin-exchange field $b_S^K S_+$ (where $b_S^K$ is the spin-exchange coefficient characterizing the influence of the Rb polarization on the Xe)~\cite{Korver2015}. This suppresses $K_z$ such that $K_+$ is much more sensitive to $\Omega_z^K$ than $\Omega_+^K$. Assuming our transverse optical pumping produces negligible $S_z$, the Xe resonance frequencies are then $\Omega_z^K=\gamma^K(B_{z0}+B_p)\mp\omega^R$, where $B_{z0}$ is the stray field inside our magnetic shields and $B_p$ is the field from the bias pulses. Since the Xe precess on the order of only $2\pi/10^3$ radians per pulse, the pulsed field can be approximated as a continuous field with $B_p= \omega_p/\gamma^S$ for pulsing frequency $\omega_p$.

The transverse Rb polarization ($S_+$) is used to polarize the Xe via spin-exchange optical pumping, and the longitudinal Rb polarization ($S_z$) is used to detect the Xe precession. For small Rb precession ($\Omega^S<< \Gamma'$), the time-average solution to the Bloch equation for the Rb polarization can be expanded~as
%Is it bold necassary? if not please delete it through the whole text.
\begin{equation} \label{Seq}
\mathbf{S} = {\mathbf{R}\over \Gamma'} +{\mathbf{\Omega}^S \times \mathbf{R} \over\Gamma'^2}+{\mathbf{\Omega}^S \times \mathbf{\Omega}^S \times \mathbf{R}\over\Gamma'^3}+..., 
\end{equation}
where $\mathbf{R}= \int d\nu \Phi(\nu) \sigma(\nu) \mathbf{p} /A$ is the pumping rate, $\sigma(\nu)$ is the cross section of the atomic transition, $\int d\nu \Phi(\nu)=P/h\nu$ is the total photon flux for a beam of power P incident on area A, $\mathbf{p}$ is the photon spin, $\Gamma'$ is the total relaxation rate (including pumping), and $\mathbf{\Omega}^S = \gamma^S (\mathbf{B}_0+b_a^S \mathbf{K}^a+b_b^S \mathbf{K}^b)$. The~magnetic field experienced by the Rb includes the stray fields $\mathbf{B}_0$ and the spin-exchange fields ($b_K^S$ is the spin-exchange coefficient characterizing the influence of the Xe polarization on the Rb), but it does not include the bias~field. This is because the bias field is applied as a sequence of low duty cycle pulses, the~area of which correspond to $2\pi$ precession of the Rb atom. These equations assume negligible back polarization from the Xe to the Rb~\cite{Limes2018}, and that $\mathbf{K}$ precesses slowly such that $S_z$ responds adiabatically.  We optically pump along $\hat{x}$ such that $\mathbf{R}=R(t)\hat{x}$, and the solution to Equation~(\ref{Seq}) is
\begin{subequations}  
\label{allequations} 
\begin{eqnarray}
S_+ = {R(t)\over\Gamma'}+i{R(t) \Omega_z^S\over \Gamma'^2} = {R(t) \over \Gamma'}e^{i\epsilon_z},
\\
S_z = -{R(t)\over \Gamma'^2}(\Omega_y^S -{\Omega_z^S\over \Gamma'} \Omega_x^S)={-R(t)\over \Gamma'^2 }\text{Im}[\gamma^S b_K^S K_+e^{-i\epsilon_z}],
\end{eqnarray}
\end{subequations}
where 
\begin{equation} \label{ezdef}
\epsilon_z = \tan^{-1}({ S_y \over S_x}) \equiv \tan^{-1}({ B_{z0} \over B_w})<<1
\end{equation}

is the magnetometer phase shift. $B_w=\Gamma'$/$\gamma^S$ is the magnetic width of the magnetometer. A stray $B_z$ effectively rotates the quantization axis of the Rb magnetometer thus causing a phase shift ($\epsilon_z$) in the measurement of a rotating $B_\perp$.

The Xe NMR can be driven by modulating the transverse Rb polarization near each isotope's resonance frequency while the bias pulsing frequency is kept fixed. Such an excitation scheme is desirable as it effectively AC couples $S_z$. Hence, we Fourier expand the Rb polarization as  $S_+(t) = e^{i\epsilon_z} \sum_{p,q} s_{pq} e^{i(p\omega_1+q\omega_2) t}$ where $s_{pq}$ is the Fourier coefficient of $S_+$ at $p\omega_1+q\omega_2$. With the substitution $K_{+} = K_{\perp}e^{\pm i(\omega_d t +\delta)}$, where $\delta$ is the phase shift of the Xe relative to the phase of $S_+$ and $\omega_d$ is the drive~frequency, we find the real part of Equation ~(\ref{BE1}) to be
\begin{linenomath}
\begin{equation}
{d K_{\perp}\over dt} =-\Gamma_2 K_{\perp}+\Gamma^K_S \sum_{p,q} s_{pq} \cos[(\omega_d+p\omega_1+q\omega_2)t + \delta \pm \epsilon_z].
\end{equation}
\end{linenomath}

The steady state solution of this equation is $K_{\perp} = \Gamma^K_S s_{pq} /\Gamma_2$ when $\omega_d$ is chosen to satisfy the resonance condition $\omega_d=-p_0\omega_1-q_0\omega_2$ for some $p_0$ and $q_0$ and when $\delta,\epsilon_z<<1$. Similarly, the~imaginary part of Equation~(\ref{BE1})  (once again for $\delta,\epsilon_z<<1$) is
\begin{linenomath}
\begin{equation}
{d \delta \over dt}= - \Delta - \Gamma_2 (\delta \mp \epsilon_z),
\end{equation}\label{fundgyro}
\end{linenomath}
where $\Delta =\omega_d- \gamma B_z\pm \omega^R$. The sign in front of $\epsilon_z$ is isotope dependent because the Xe isotopes precess in opposite directions. We can solve this equation in the Fourier domain to find
\begin{linenomath}
\begin{equation}
\tilde{\delta} = -{\tilde{\Delta} \mp \Gamma_2 \tilde{\epsilon}_z \over \Gamma_2+i\omega},\label{fundgyro1}
\end{equation}
\end{linenomath}
where the notation $\tilde{f} = f(\omega)$ is used.   

By monitoring $S_z$, we can obtain measurements of $\delta$ for each Xe species. These measurements of $\delta$ are used to extract the Xe resonance frequencies. We perform comagnetometry using these Xe resonance frequencies to remove magnetic field correlations and arrive at a measurement of inertial~rotation. In~Section~\ref{apparatus} we detail the implementation of our Xe excitation scheme, and in Section~\ref{result} we demonstrate the performance of our system.

\section{Materials and Methods} \label{apparatus}
\begin{figure}[h]%%
\centering
\includegraphics[width =0.8\linewidth]{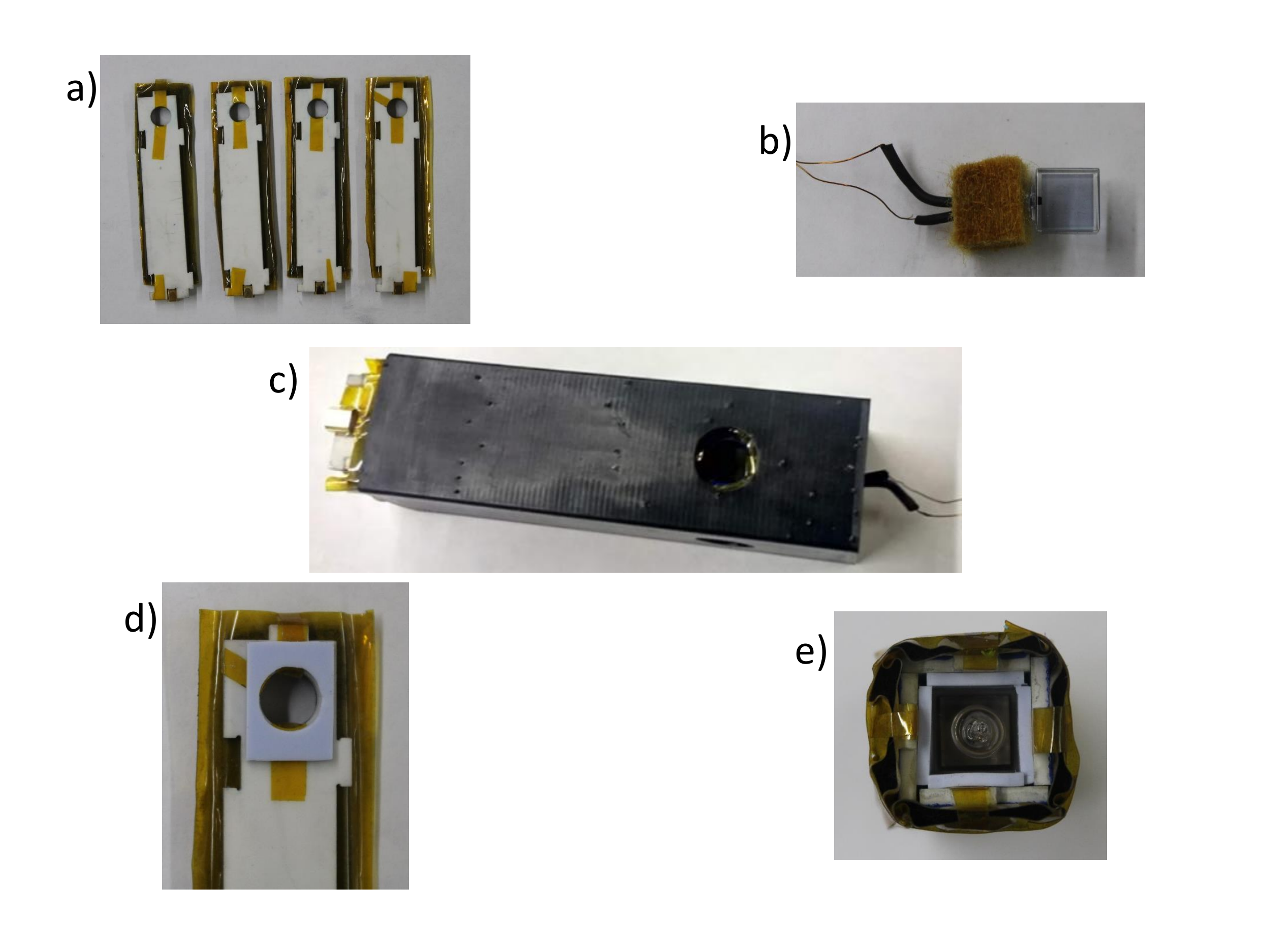}
\caption[Vapor cell heater and cartridge mount photographs.]{Vapor cell heater and cartridge mount photographs. (\textbf{a}) Four ceramic jig heater sides with aerogel pillows attached to outer faces. AC power enters via MMCX connectors at the bottom of each jig heater. (\textbf{b}) 1 cm$^3$ vapor cell with stem tucked into fiberglass insulation. A temperature sensor sits between the insulation and outer cell wall. (\textbf{c}) Cartridge with cell installed. (\textbf{d}) Gap fill shim on one jig heater face. (\textbf{e}) View looking into the jig heater with the cell installed. Note the four pieces of gap fill between each outer cell wall and the jig heater inner faces.}\label{cellpics}
\end{figure}

An 8 mm inner diameter cubic Pyrex cell filled with 68 Torr enriched Xe and 85 Torr N$_2$ with a~hydride coating~\cite{Kwon1981} is mounted in a ceramic housing with holes for laser light to enter the cell. The~ceramic housing has four symmetric faces which fit together like a jigsaw puzzle (and so we call these ``jig'' heaters). On each face are printed conductive traces through which AC current at $\sim$150 kHz is passed to heat the ceramic. The conductive traces are arranged to produce minimal stray magnetic fields including gradient magnetic fields. The $\sim$1 mm gap between the vapor cell and ceramic heating jig is shimmed with a 1.5 mm thermally conductive and slightly compressible gap fill (model: TG~977, manufacturer: T-Global Technology, see Figure~\ref{cellpics}). The ceramic is wrapped with aerogel (a high temperature insulating material) and secured with Kapton tape and then fitted into a 3D printed (high~temperature nylon) cartridge with holes to allow laser light to enter the cell. The~compressible nature of the aerogel produces a friction fit keeping the ceramic jig structure fixed within the cartridge. The cartridge itself is mounted in a 3D printed (ABS plastic) rig. This rig has support arms which extend out three of the magnetic shield portholes described below. These support arms are secured directly to an optical table on which the entire apparatus is mounted.
\vspace{-6pt}

The three layer $\mu$-magnetic shield we use is cylindrical with eight access ports (two along the axis of symmetry and six oriented tangentially). Since the tangential access ports are not located halfway between the two ends of the cylindrical shields, and since pump laser light must enter through the~ports, we must place our cell such that it is not equidistant from the end caps. The asymmetry in distance to end caps informs the design of our bias pulsing coil set. To minimize coupling to the shield end caps and maximize uniformity across the volume of the cell, the pulsing coil set consists of two pairs of square Helmholtz coils with differing side lengths wound in series with opposite~polarity. The purpose of the ancillary counter wound coils is to suppress the field produced by the coil set at the nearest end cap. See \cite{KorverThesis} for specific design details.

The bias field requires short pulses ($<$5 $\upmu$sec) of $\sim$1 Ampere peak current. The circuit used to drive the pulsing coil was custom-made and is described in \cite{KorverThesis}. The circuit used to drive the shim coils was also custom-made and is described in \cite{WyllieThesis}.

\begin{figure}[h]%%
\centering
\includegraphics[width =1\linewidth]{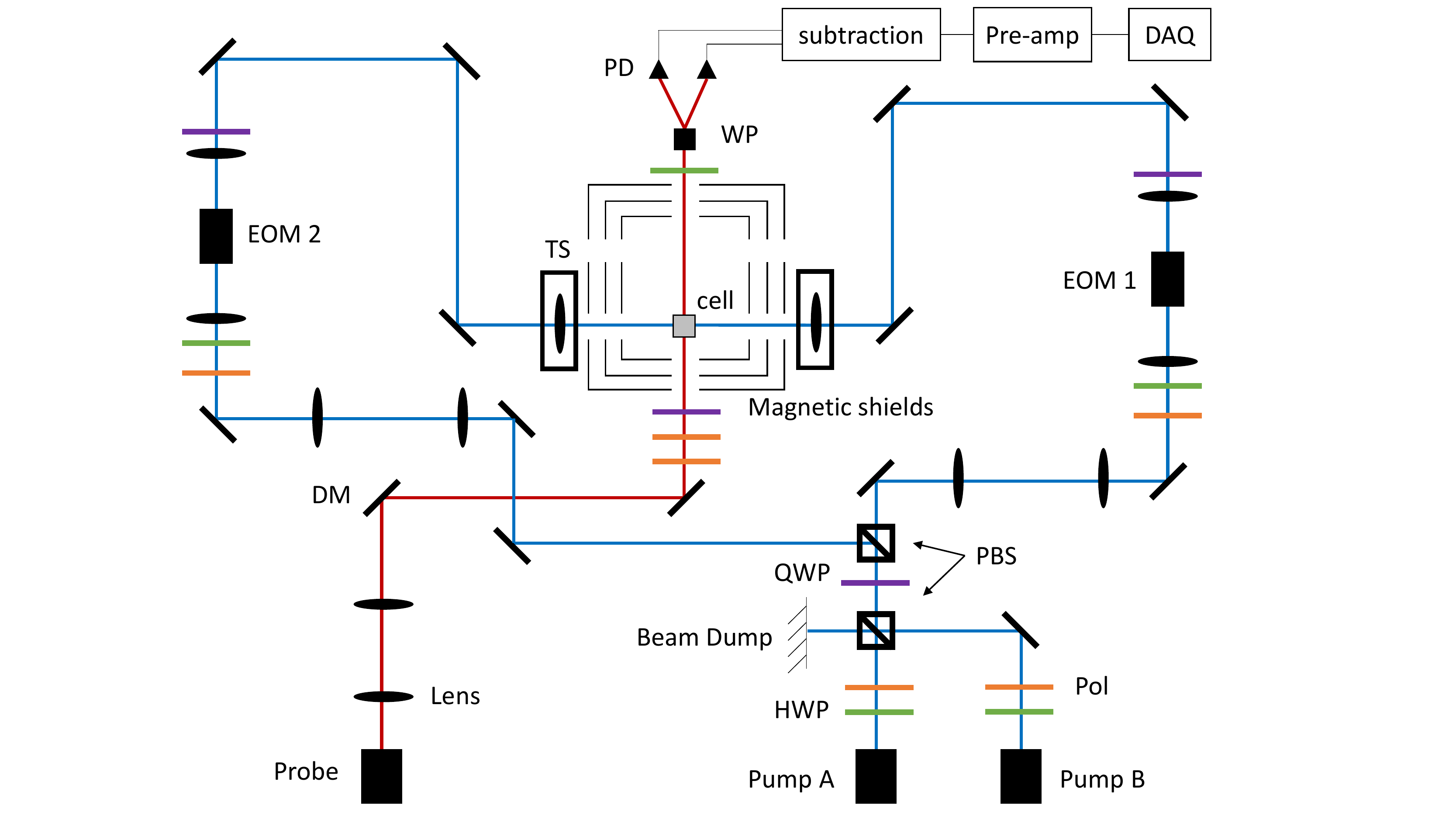}
\caption[Experimental setup for PM excitation]{Experimental setup for PM excitation (not to scale). DM: dichroic mirror, Pol: polarizer, HWP: half wave plate, QWP: quarter wave plate, PBS: polarizing beam splitter, WP: Wollaston prism, PD:~photodiode, EOM: electro-optic modulator, TS: two-axis translation stage with lens. The three-axis magnetic shim and pulsing coils are not shown. The setup fits on a four foot square optical table.}\label{polmodsetup}
\end{figure}

To perform optical pumping of the Rb, the outputs of two distributed feedback laser diodes tuned near the Rb D1 transition (one on either side of resonance) are overlapped (see Figure~\ref{polmodsetup}). This is accomplished by polarizing pump A so that it is mostly transmitted by a polarizing beam splitter (PBS) and pump B so that it is mostly reflected by the same PBS. The combined beam is then sent through a quarter wave plate and then separated into two beams using a PBS once again. The orientation of the quarter wave plate is chosen such that both pump A and pump B have half their power in each output beam. A telescope is used to couple the beams into individual EOMs. Prior to each EOM is a polarizer and half wave plate. The polarizer ensures that the light incident to the EOM crystal is purely~linear. The half wave plate is used to align the light polarization relative to the EOM crystal~axis. The maximum and minimum voltages of the EOM drive waveform are chosen to produce $\pm\lambda/2$ retardance. The quarter wave plate at the output of the EOM converts the EOM output at $V_{max}(V_{min})$ to be $\sigma^+(\sigma^-)$.  The collimated output of each EOM is coupled into the vapor cell from opposing directions. The overall beam resizing is set to somewhat overfill the aperture of the ceramic~heater. Fine tuning of each pump beam's pointing is controlled using long focal length lenses mounted on two axis translation stages just outside the magnetic shield. The position of each steering lens is chosen to optimize the magnetometer gain. The power and detuning of each pump laser is chosen to approximately cancel the AC Stark effect. While we could also suppress the AC Stark effect by pumping with a single laser tuned on resonance, we would not get good spatial uniformity of the Rb polarization without significantly increasing our laser power due to optical thickness effects. To~detect $S_z$, approximately one mW of linearly polarized light from the output of a third distributed feedback laser diode, tuned near the Rb D2 line, is directed through the center of the cell and parallel to $\hat{z}$ onto a balanced Faraday detector.

We choose to modulate the Rb polarization by switching between just two polarization states, $\sigma^{\pm}$. We modulate the x-component of $\mathbf{R}$ according to
\begin{equation}
R(t) = R_{0}\;\text{sign}[\cos\left({\omega_d^a+\omega_d^b\over 2}t+2\cos({\omega_d^a-\omega_d^b\over 2}t)\right)].
\end{equation}

This waveform is advantageous
%unique 
in that the smallest separation between reversals is larger than the finite response times of the EOMs and of the optical pumping of the Rb atoms. Modulating the polarization as a sum of two sine waves would also keep the modulations within the bandwidth of the EOMs and the Rb magnetometer, but we find that producing such a PM waveform in our optically thick vapor cell is very challenging. We use the same modulated cosine waveform to apply a~compensation field, the amplitude of which is set to cancel the spin-exchange field experienced by the Xe from the Rb. Doing so suppresses the production of $K_z$ and narrows the NMR linewidths, $\Gamma_2$~\cite{Korver2015}.

From Equation~(\ref{allequations}) we find that the detected longitudinal Rb polarization $S_z$ is 
\begin{equation}\label{Szdet1}
S_z = -{R(t)\over \Gamma'^2}[b_a^S K^a_{\perp} \sin(\delta^{a}+\alpha^a - \epsilon_z)+ b_b^S K^b_{\perp} \sin(\delta^{b}+\alpha^b + \epsilon_z)],
\end{equation}
where $\delta=\phi-\alpha$ is the difference between the drive phase $\alpha=\int dt \omega_d$ and the Xe precession phase $\phi$ for each isotope,  $\epsilon_z$ is the magnetometer phase shift, and $R\sim S_+$ is the optical pumping rate of the Rb. Although lock-in detection can be accomplished on $S_z$ as it appears in Eq.~\ref{Szdet1}, the phase sensitivity is diminished due to the presence of $R(t)$ which effectively mixes some Xe signal to DC. A more effective approach is to "rectify" the $S_z$ signal such that $R(t)$ is removed. This is accomplished by multiplying $S_z$ by $R^{-1}(t)$. The resulting signal is then sent into two separate lock-in amplifiers, each referenced to a different isotope's drive frequency. From these demodulations, we arrive at the measured phases $\delta^a-\epsilon_z$ and $\delta^b+\epsilon_z$.

\section{Results} \label{result}
\vspace{-6pt}
\subsection{NMR Excitation and Detection}

\begin{figure}[t]%%
\centering
\includegraphics[width=0.75\linewidth]{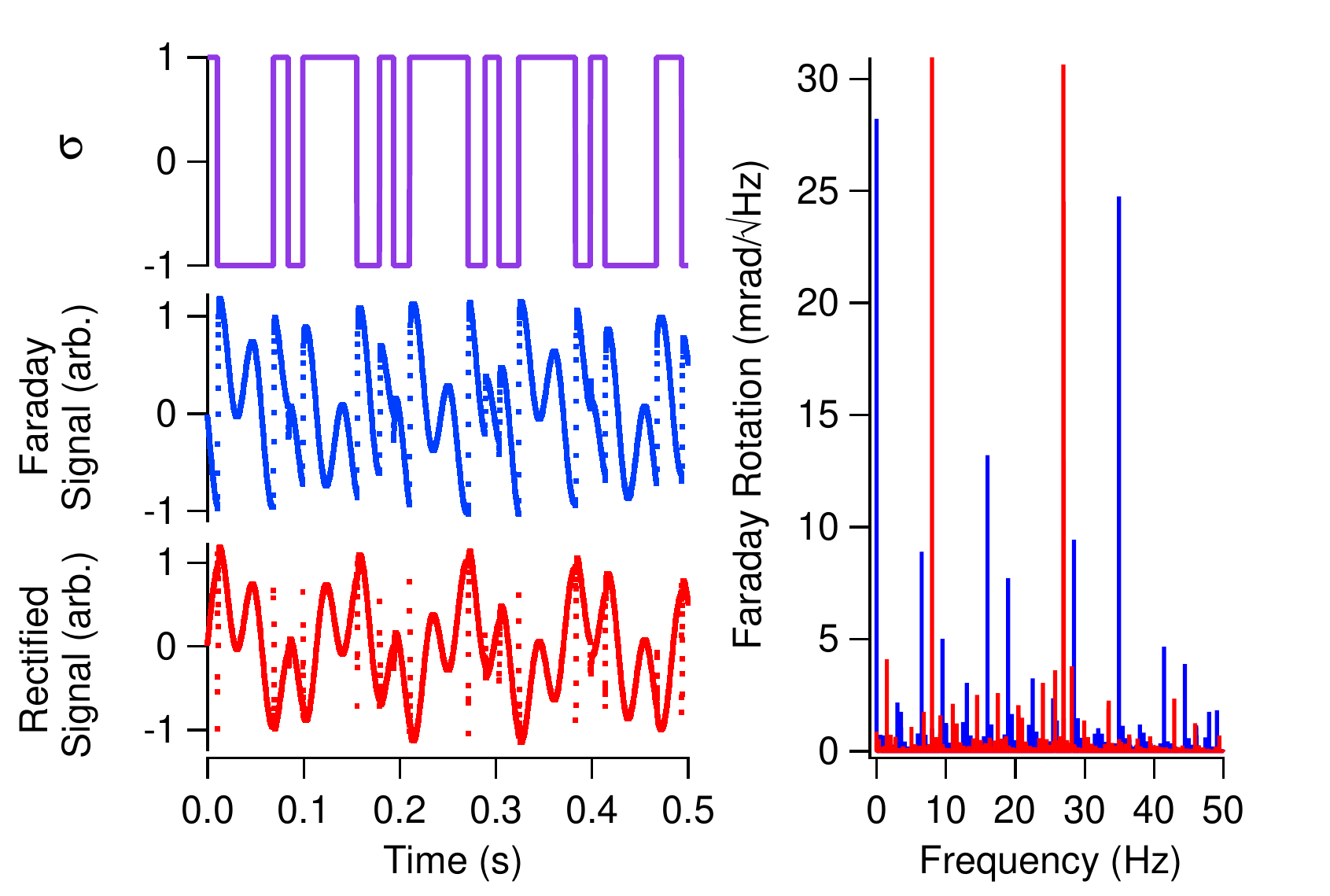}
\caption[Square wave PM signals]{Measured square wave PM signals. Top: $S_+(t)$. Middle: $S_z(t)$. Bottom: Rectified $S_z(t)$. The plot on the right shows the amplitude spectral density with (red) and without (blue) rectification.}\label{polmodsqsigs}
%\end{figure}
%\unskip
%\begin{figure}[h]%%
%\centering

\vspace*{\floatsep}

\includegraphics[width=0.75\linewidth]{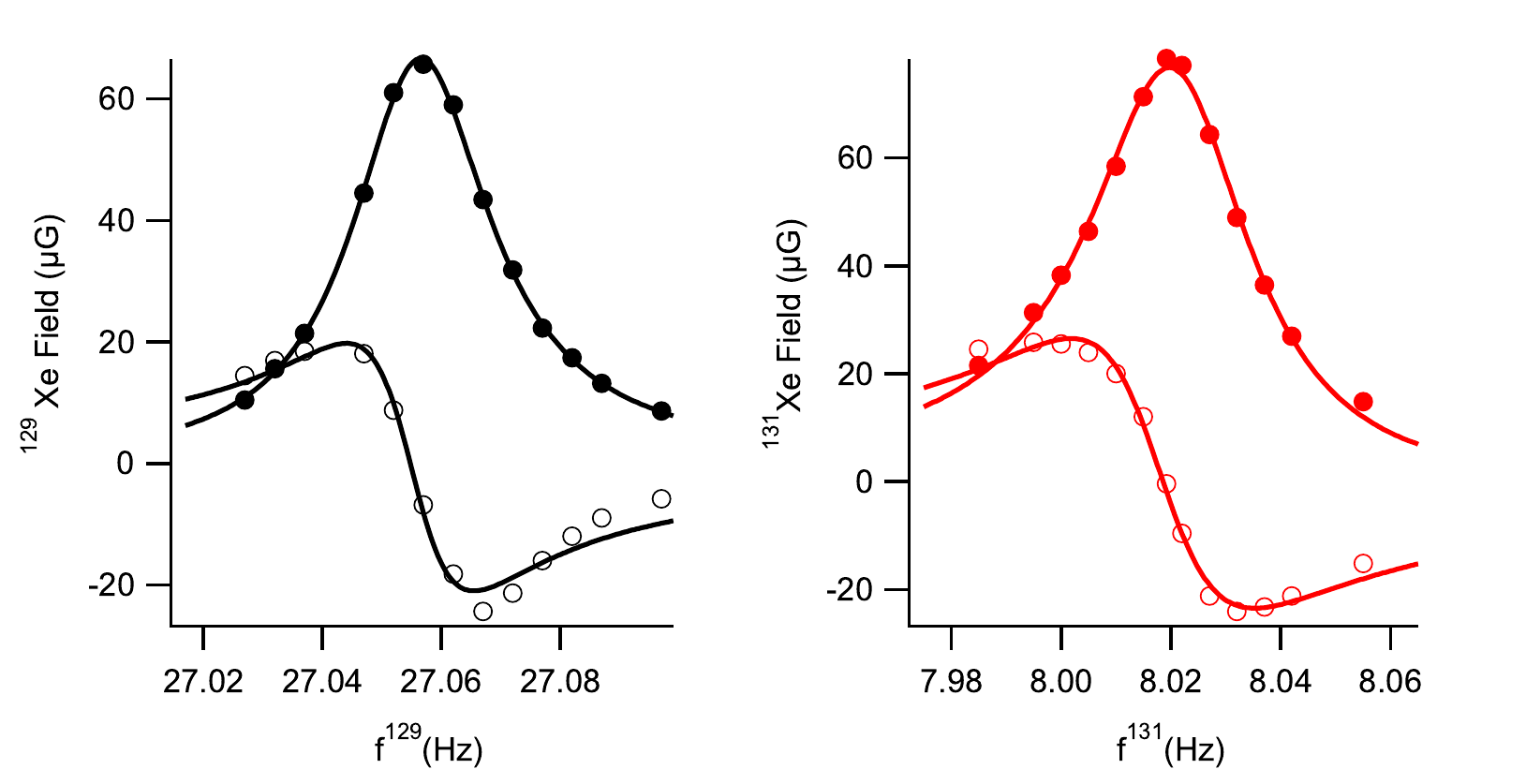}
\caption[NMR of each species]{Measured NMR lineshapes of each species. Filled circles are $K_x$, open circles are $K_y$, and lines are Lorentzian fits.}\label{polmodnmr}
\end{figure}

Figure~\ref{polmodsqsigs} shows the time series of the measured $S_z$ with and without rectification when we drive both isotopes near resonance simultaneously using $S_+$. We see that rectification reveals the sinusoidal precession of each isotope. The outlying data on the rectified signal, which occur when the polarization is reversed, are due to the finite response time of the Rb magnetometer. Although rectification collects the many Xe signal sidebands into the two Xe carrier frequencies (see power spectrum on right), it~also maps 1/$f$ detection noise on $S_{z}$ to the carrier frequencies. The mapping of low-frequency $S_{z}$ noise to the carrier frequencies can be prevented by high-pass filtering $S_z$ with a 1 Hz corner prior to rectification. 

Figure~\ref{polmodnmr} shows the detected NMR signals for each isotope. These data were acquired by driving one isotope on resonance while varying the other isotope's drive frequency and recording its $K_x$ and $K_y$ derived using demodulation. We see that the lineshapes are nearly Lorentzian with linewidths (half-width at half-max) of $\sim$15 mHz and amplitudes of approximately 60 $\upmu$G. The on-resonance amplitudes are in agreement with estimates similar to those outlined in~\cite{Walker2016}. The implied $T_2^a$, $T_2^b$ from the fits are in good agreement with independent measurements of each isotope's $T_1$. The $\sim$15 mHz linewidths are only possible because of two features of our experiment; (i) the use of a Rb hydride cell coating (without which $T_2^b$ would be substantially shorter and $T_2^a$ much longer)~\cite{Kwon1981}, and (ii) the application of a magnetic compensation field $B_x$ that cancels the Rb spin-exchange field experienced by the Xe. 

\begin{figure}[h]%%
\centering
\includegraphics[width=0.75\linewidth]{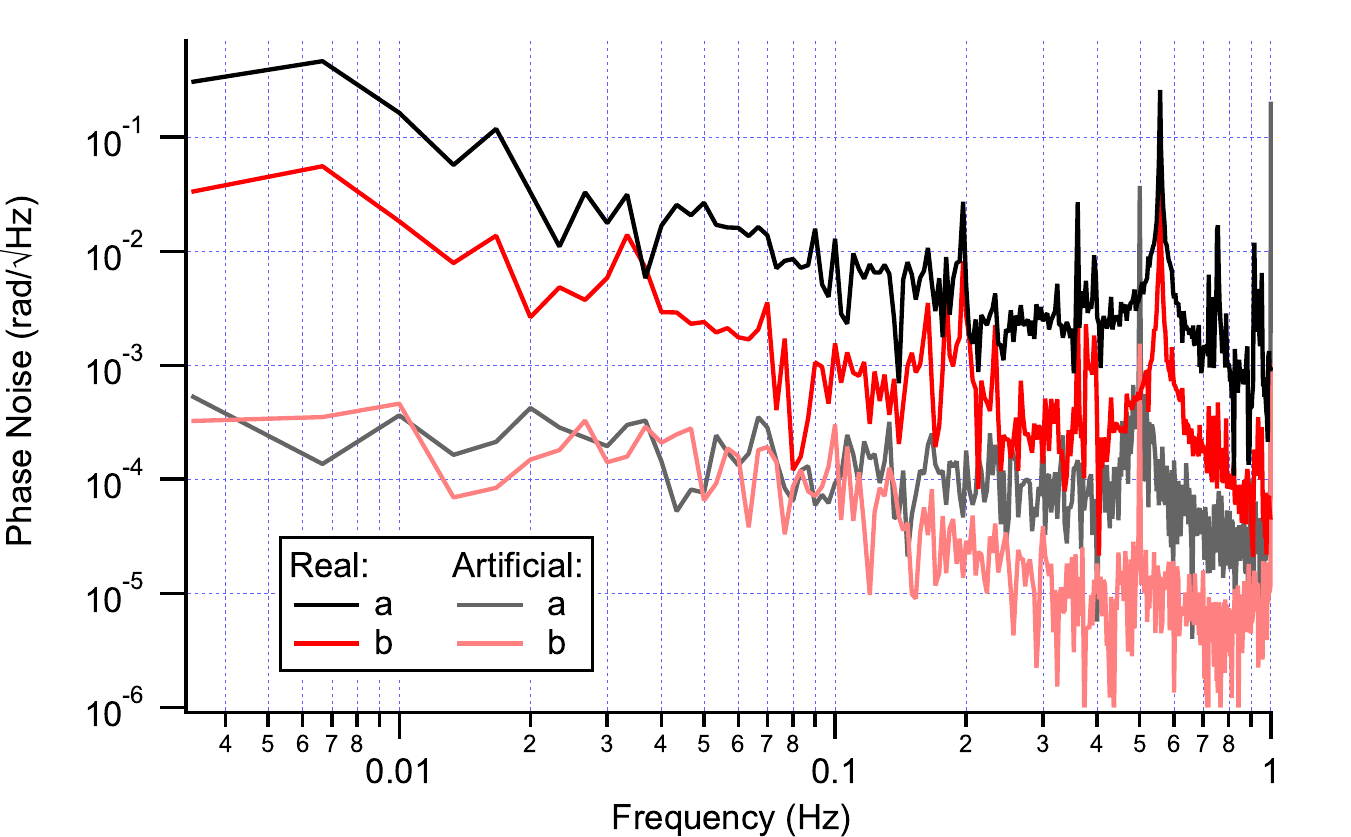}
\caption[Phase noise]{Measured phase noise of each species. Traces labeled ``Real'' are recorded when the Xe isotopes are excited on resonance. Traces labeled ``Artificial'' are recorded when the Xe isotopes are both driven off resonance (not excited) and an AC $B_y$ is applied to the magnetometer at the off-resonance drive frequencies.}\label{polmodpn}
\end{figure}

Figure~\ref{polmodpn} demonstrates the amplitude spectral density of the phase noise measured for each isotope under simultaneous resonant excitation conditions. We see that for frequencies less than 1 Hz the spectra are dominated by $1/f$ noise which is about $\rho=\gamma^a/\gamma^b$ greater for isotope $a$ (black traces) than for isotope $b$ (red traces) suggesting the dominant source of $1/f$ noise is magnetic in~nature. Also shown is the phase noise measured when the Xe isotopes are driven off resonance (not excited) and a so-called ``artificial'' Xe signal is applied as an AC field, $B_y$. The amplitudes of this signal $A^a\sin(\omega_d^a t)+A^b\sin(\omega_d^b t)$ are chosen to produce the same size magnetometer signal as real Xe. We~note that this artificial Xe signal is planar, unlike the real Xe signal which rotates. The artificial signal allows us to measure the signal-to-noise ratio (SNR) of the Rb magnetometer. These signals do not show $1/f$ dependence because, unlike the real Xe phase, the SNR of the magnetometer does not depend on the bias magnetic field to first order. The detection phase noise from each artificial Xe measurement is uncorrelated and limits the possible field suppression when performing comagnetometry. The SNR for each isotope is $\sim$5000$\sqrt{\text{Hz}}$. While the detection of artificial Xe is insensitive to $1/f$ bias magnetic field noise, $1/f$ $S_z$ noise can still be mapped to the artificial Xe frequencies via rectification, and so the addition of a 1 Hz high-pass filter prior to rectification was essential for realizing such an SNR.

\subsection{Comagnetometry}

\begin{figure}[h]%%
\centering
\includegraphics[width=0.70\linewidth]{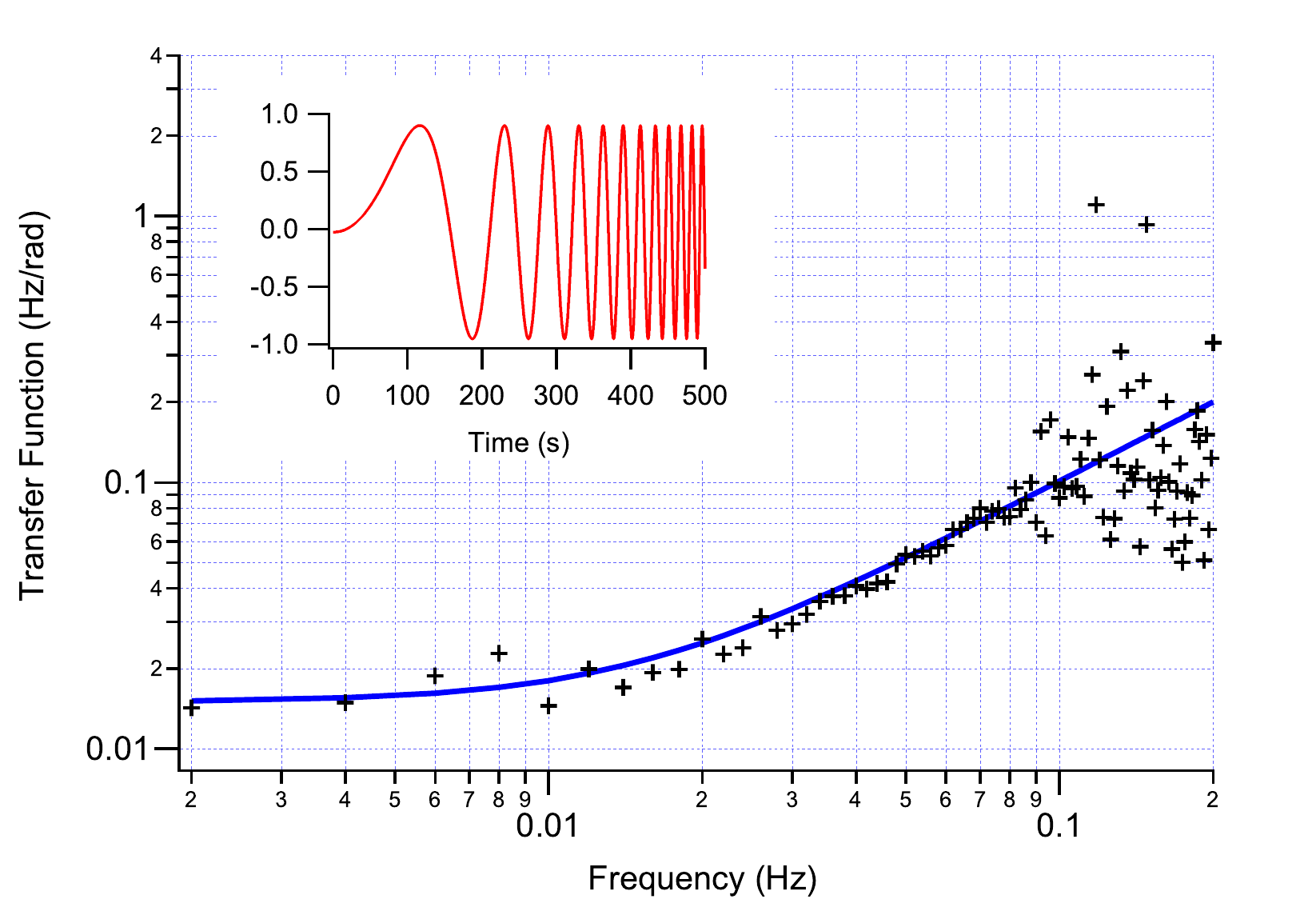}
\caption[Transfer function of $^{129}$Xe]{Transfer function of $^{129}$Xe. Inset depicts the normalized chirp waveform used to modulate the bias field. Black crosses are measured data, while the blue line is a fit of the form $\sqrt{(\Gamma_2^{a})^2+f^2}$.}\label{polmodtf}
\end{figure}

We perform comagnetometry by subtracting magnetic field correlations between the two isotope's precession frequencies. Since our device measures phase, we need to know or measure the transfer function from phase to frequency. In Section~\ref{BEsec} we derived the transfer function (see Equation~(\ref{fundgyro1})). We measured the transfer function of each isotope by recording the response of the measured phase $\delta$ to sinusoidal modulation of $B_z$. Figure~\ref{polmodtf} shows the measured transfer function for isotope $a$. We~use a~chirp waveform to modulate the bias field $B^{mod}_z(t)=B_0\sin(2\pi[e^{t/T_2}-1 - t/T_2])$ (where~$T_2 = 1/2\pi \Gamma_2$), the time series of which is shown in the inset of Figure~\ref{polmodtf}. This modulation waveform allows us to measure the transfer function from 0.002 to 0.1 Hz with good SNR in a single data acquisition. The~transfer function is the ratio $\gamma^K\tilde{B}_z^{mod}/\tilde{\delta}$. We fit the data according to Equation~(\ref{fundgyro1}) and find excellent agreement with the linewidth derived from the fits in Figure~\ref{polmodnmr}.

\vspace{-6pt}
\unskip

Although conversion from phase to frequency for the measured Xe phases is possible using a~measured transfer function, feedback is desirable because (in the high gain limit) the performance of the comagnetometry becomes insensitive to changes in the transfer function. We used the measured precession phase of isotope $a$ to stabilize the bias field and the measured transfer function of isotope $b$~to convert its measured phase noise to frequency noise. Under such conditions the frequency noise of isotope $b$ is proportional to rotation. We apply a feedback field, $B_f$, to hold the measured phase of isotope $a$, $(\delta^a-\epsilon_z)$, equal to zero. The field stabilization can then be written $\tilde{B}_f = \tilde{G}(\tilde{\delta}^a-\tilde{\epsilon}_z)$. In the high gain limit, this becomes
\begin{linenomath}
\begin{equation}
\lim_{G \to \infty}\tilde{B}_{tot} = {1\over \gamma^a}(\tilde{\omega}^R+i\omega\tilde{\epsilon}_z),
\end{equation}
\end{linenomath}
where $B_{tot}=B_{z0}+B_p+B_f$, and where we made substitutions using Equation~(\ref{fundgyro1}) and $\tilde{\Delta}^a =- \gamma^a \tilde{B}_z- \tilde{\omega}^R$ (since $\omega_d^a$ is held constant). The measured phase of isotope $b$, $(\delta^b+\epsilon_z)$, converted to frequency is
\begin{linenomath}
\begin{equation}
\tilde{\omega}^b \equiv (\tilde{\delta}^b+\tilde{\epsilon}_z)(\Gamma_2^b+i\omega)= (\rho^{-1}+1)(\tilde{\omega}^R+i\omega\tilde{\epsilon}_z).
\end{equation}
\end{linenomath}

We see that in the high gain limit, when correcting the bias field to keep the measured phase of isotope $a$ equal to zero, the rotation is simply $\tilde{\omega}^R = \rho\;\tilde{\omega}^b/(1+\rho)$ assuming $\tilde{\epsilon}_z$ is negligible.

\begin{figure}[h]%%
\centering
\includegraphics[width=0.9\linewidth]{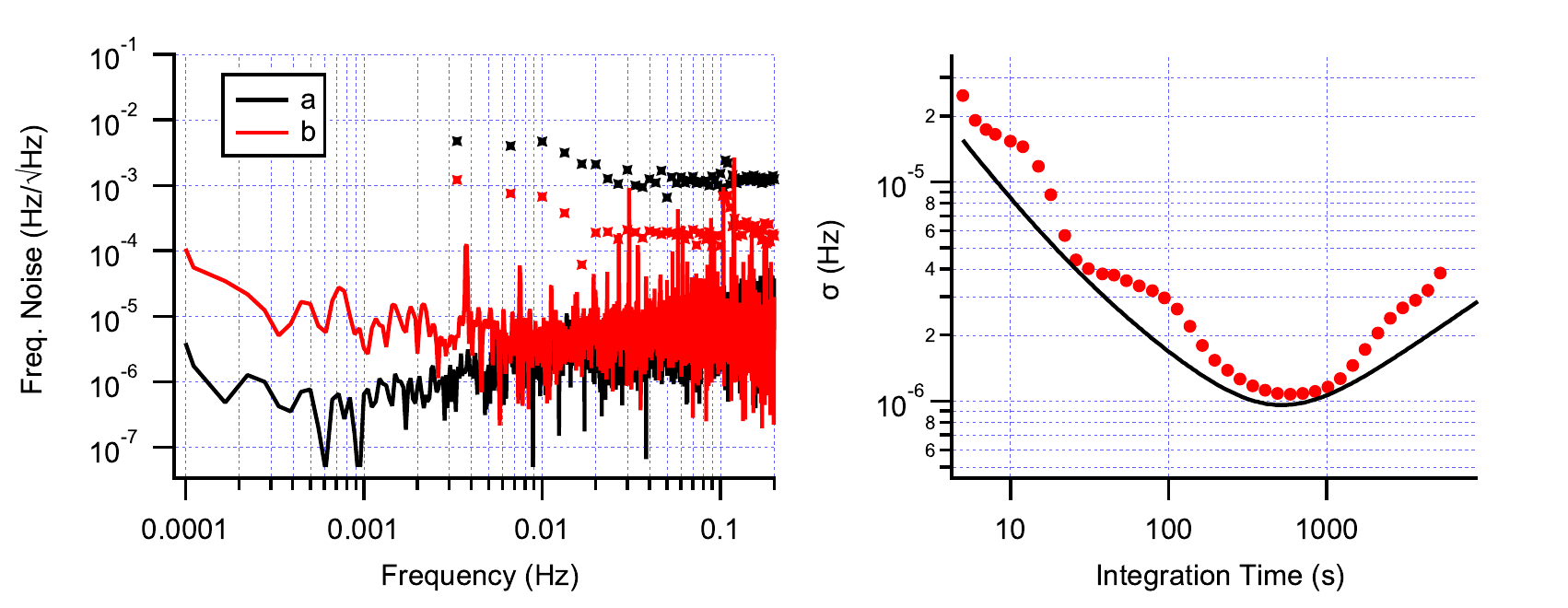}
\caption[Comagnetometry noise and stability]{Measured comagnetometry noise and stability. Left: the amplitude spectral density of frequency noise. The cross marks indicate open loop frequency noise. The solid lines are frequency noise when the measured phase noise of isotope $a$ is used to stabilize bias field. Right: modified Allan deviation of $\omega^b$. Filled circles are measured data. Solid line shows the quadrature sum of $7\times 10^{-5}\tau^{-1}$, $15\;\upmu\text{Hz}/\sqrt{\text{Hz}}\tau^{-1/2}$, and $30\;\text{nHz}\sqrt{\text{Hz}}\tau^{1/2}$ trends.}\label{polmodcomag}
\end{figure}

The best performance we observed with bias field feedback activated is shown in Figure~\ref{polmodcomag}. The~feedback consisted of two analog inverted zero gain stages. The influence of bias field feedback is dramatic from 0.1 to 200 mHz. The servo suppresses $\tilde{\omega}^a$ to below 1 $\upmu$Hz/$\sqrt{\text{Hz}}$ at low frequency, which is nearly $10^4$x less than the open loop noise. Because magnetic noise dominates each isotope's precession, servoing the measured phase of isotope $a$ also greatly suppresses $\tilde{\omega}^b$. We observe at least 100$\times$ improvement in $\tilde{\omega}^b$ due to feedback. The modified Allan deviation~\cite{Allan1981} suggests a rotation ARW sensitivity of $\sqrt{2}\;15\;\upmu$Hz/$\sqrt{\text{Hz}}{\rho\over 1+\rho}\sim 16\; \upmu$Hz/$\sqrt{\text{Hz}}$ and a rotation bias instability of 1 $\upmu$Hz ${\rho\over 1+\rho}\sim 800$~nHz. The size of ARW is within a factor of 3 of the ratio of the measured linewidths divided by the SNRs (shown in Figure~\ref{polmodpn}). The peaks in the Allan deviation at 15 and 100 s of integration are due to low-frequency narrow-band large-amplitude noise peaks in $\tilde{\omega}^b$ which we attribute to the PM~waveform. The bias instability is limited by $\tau^{1/2}$ trending noise of unknown origin. We find that feedback causes the measured phase of isotope $b$ to trend linearly in time. The source of this frequency bias is uncertain. Although the trend is very stable over the course of a data run, the trend is not consistent between data runs. The bias instability demonstrated in Figure~\ref{polmodcomag} was difficult to reproduce. Typically, the bias instability we measured was a few $\upmu$Hz.

\subsection{Cross Talk}
Once we measured the stability of the PM comagnetometer, we desired to know the fidelity with which our detection separated signals from the two Xe isotopes. It is possible that the detection channel designed to measure isotope $a$’s phase was really measuring a linear combination of isotope $a$~and $b$’s phases. If such ``cross talk'' were present then the scale factor (or how we convert the measured precession frequencies to rotation) would change~\cite{ThrasherThesis}. Suppose there exists cross talk in both channels such that $\omega^a = \gamma^a B_z + \beta \omega^b - \omega^R$ and $\omega^b = \gamma^b B_z +\beta' \omega^a + \omega^R$ where $\beta$ and $\beta'$ represent the cross talk between detection channels. Solving for $\omega^R$ we find
\begin{linenomath}
\begin{equation}
\omega^R = {\omega^b (\rho+\beta')- \omega^a (\rho \beta +1) \over 1+\rho},
\end{equation} \label{polmodcteq}
\end{linenomath}
where if $\beta=\beta'=0$ we return to the expected expression for rotation without cross talk. Non-zero cross talk is undesirable because the accuracy with which it is known (or measured) limits the accuracy of conversion from measured precession frequencies to rotation (or any other non-magnetic spin-dependent interaction). A measurement of cross talk is vital since an important alleged feature of our comagnetometer is that it has a scale factor which is determined solely by $\rho$.

\begin{figure}[h]%%
\centering
\includegraphics[width=0.65\linewidth]{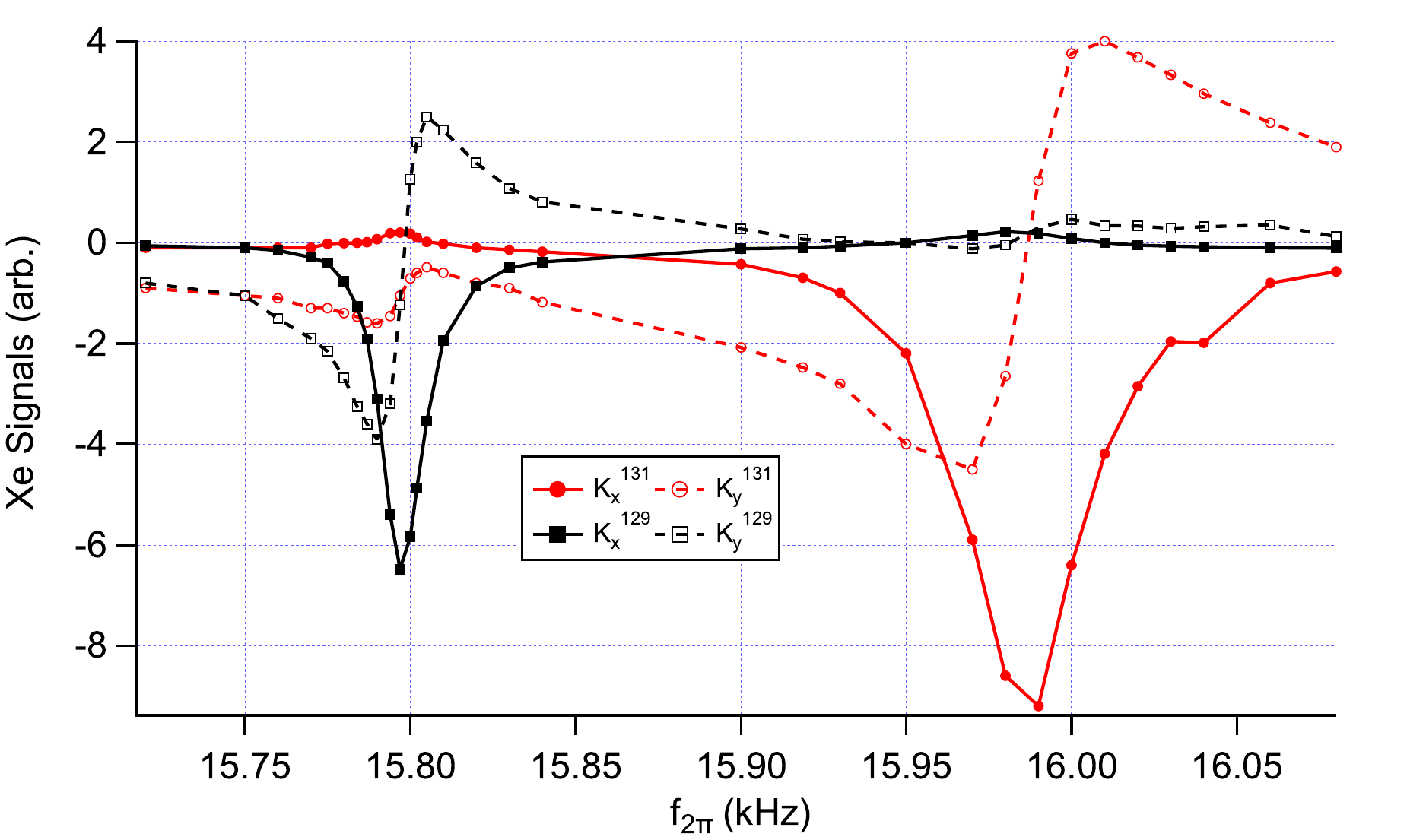}
\caption[Measurement of cross talk]{Measurement of cross talk for detection with rectification. The Xe drive frequencies are set such that the two isotopes are not driven on resonance at the same effective bias field magnitude. The~in-phase (filled symbols) and out-of-phase (un-filled symbols) components of each isotope's detection channel (black and red correspond to isotopes $a$ and $b$, respectively) are shown as the bias pulsing frequency is varied. We see that when isotope $a$ is resonant ($f_{2\pi}\sim 15.8$ kHz) the signal on isotope $b$’s detection channel is not flat despite isotope $b$ being driven many linewidths off resonance. Similarly, when isotope $b$ is resonant ($f_{2\pi}\sim 16$ kHz) the signal on isotope $a$’s detection channel is not flat despite isotope $a$ being driven many linewidths off resonance.}\label{polmodmct}
\end{figure}

We characterize the cross talk present in our detection channels by looking for changes in the detected signal for one isotope when the drive of the other isotope is changed. We do this by detuning one isotope’s drive frequency by $\sim300$ mHz and then scanning the bias pulse repetition~rate. The~effective resonance frequency of each Xe isotope depends on the pulsing frequency ($\omega_{2\pi}$) as $\omega^K = \omega_{2\pi}\gamma^K/\gamma^S$. The 300 mHz detuning ensures that both isotopes are not simultaneously on resonance for a given pulsing frequency. When a Xe isotope is far from resonance, we expect the measured signals for that isotope's detection channel to be relatively flat if there is no cross talk. Cross~talk manifests if the signals from the off-resonanct isotope are not flat when the other isotope is scanned through resonance. Figure~\ref{polmodmct} shows a phase sensitive measurement of cross talk. When~isotope $a$ is excited and isotope $b$ is not, isotope $b$'s detection channel exhibits non-zero signal, and vice versa. We estimate $\beta = Q_{pp}^b/Q_{pp}^a = 0.17$ (when isotope $a$ is on resonance) and  $\beta' =Q^a_{pp}/Q^b_{pp}=0.07$ (when isotope $b$ is on resonance), where $Q^K_{pp}$ is the peak-to-peak quadrature signal of isotope $K$.

We observe cross talk on our detection channels even when Xe is not excited and artificial Xe signals are applied, suggesting that our observed cross talk is not due to physical interactions between Xe isotopes. We believe the measured cross talk stems from imperfect rectification of $S_z$ due to optical pumping transients (i.e., gain reversals) and unaccounted-for phase shifts from high-pass filtering prior to rectifying. The optical pumping transients stem from the few ms finite response time of the magnetometer. Indeed, by implementing a sample-and-hold algorithm to ignore data acquired during optical pumping transients, the cross talk is suppressed. The cross talk is further suppressed when the high-pass filter is removed. Doing so, however, reduces the detection SNR by an order of magnitude since rectification maps $1/f$ $S_z$ noise to the Xe carrier frequencies. 

\section{Discussion}

We demonstrated a novel spin-exchange pumped $^{131}$Xe-$^{129}$Xe NMR gyroscope. The~production of longitudinal spin-exchange fields and the systematic uncertainty inherent to them is greatly suppressed by synchronous spin-exchange optical pumping. The simultaneous precession of each transversely polarized noble gas is continuously monitored using the optically pumped Rb~atoms. The measured Larmor resonance frequencies are highly correlated with bias magnetic field fluctuations. 

Computing the stationary sensor's perceived inertial rotation when oriented East-West and assuming a scale factor determined solely by $\rho$ allows for an ARW sensitivity of $\sim$16\; $\upmu$Hz/$\sqrt{\text{Hz}}$ and bias instability of $\sim$800 nHz. The finite short term sensitivity appears limited by the detection~SNR. What limits the SNR is not known. The photon shot noise limited magnetometer performance should support an SNR more than 100x greater than we measure. It is possible that fluctuations in pump frequency and intensity could limit our SNR by producing noise in the AC Stark shift and the Rb polarization. We confirmed that the fidelity of modulation produced by the EOMs is sufficient to support 4x the measured SNR. The source of drift which limits the bias instability is also unknown. Cell~temperature, which was not stabilized during these measurements, could be an important contribution to long term stability. We see no direct indications of first order quadrupole from $^{131}$Xe interactions with electric field gradients at the cell walls. We showed that cross talk exists between phase sensitive detection channels for each noble gas. Such cross talk influences the device's effective scale factor. Future work will include detecting $S_z$ in a way such that gain modulation is suppressed as demonstrated in~\cite{Thrasher2019,thrasher2019PRApp}. This device's scale factor can be verified by orienting its sensitive axis along North-South and measuring Earth's rate of rotation.

%%%%%%%%%%%%%%%%%%%%%%%%%%%%%%%%%%%%%%%%%%
\begin{acknowledgments}
We would like to thank Michael Larsen for insightful discussions.

This research was funded by the National Science Foundation grant numbers PHY-1607439 and PHY-1912543 and Northrop Grumman Mission Systems' University Research Program.

All authors contributed equally in all respects to this manuscript. All authors have read and agreed to the published version of the manuscript.

The authors declare no conflict of interest. The funders had no role in the design of the study; in the collection, analyses, or interpretation of data; in the writing of the manuscript, or in the decision to publish the results.
\end{acknowledgments}

\appendix*
\section{Abbreviations}
The following abbreviations are used in this manuscript:\\
\begin{tabular}{@{}ll}
ARW & angle random walk \\
EOM & electro-optic modulator \\
GPS & global positioning system \\
NMR & nuclear magnetic resonance \\
PBS & polarizing beam splitter \\
PM & polarization modulation \\
SNR & signal-to-noise ratio \\
\end{tabular}

%Other abbreviations used in figures are defined in the figure captions.

%%%%%%%%%%%%%%%%%%%%%%%%%%%%%%%%%%%%%%%%%%

%\begin{abbreviations}
%\end{abbreviations}

\bibliography{ASv9Arxiv}

\end{document}